\documentclass[12pt]{article}
\usepackage{amsmath}
\usepackage{graphicx}
\usepackage{natbib}
\usepackage{url} 

\newcommand{\blind}{0}

\newcommand\diag{\mathrm{diag}}

\newcommand{\vat}{\mathbb{E}}

\newcommand\ind{\bot\hspace*{-6pt}\bot}  
\newcommand{\bc}{\begin{center}}
	\newcommand{\ec}{\end{center}}

\newcommand{\bit}{\begin{itemize}}
	\newcommand{\eit}{\end{itemize}}

\newcommand{\be}{\begin{eqnarray*}}
	\newcommand{\ee}{\end{eqnarray*}}
\newcommand{\ben}{\begin{eqnarray}}
	\newcommand{\een}{\end{eqnarray}}

\newcommand{\g}{\,\vert\,}
\newcommand{\gbig}{\,\Big\vert\,}

\newcommand{\D}{\mathcal{D}}

\newcommand{\N}{\mathcal{N}}

\newcommand{\X}{\mathcal{X}}

\newcommand{\pa}{\mathrm{pa}}
\newcommand{\fa}{\mathrm{fa}}

\newcommand{\bzero}{\bm{0}}

\newcommand{\bB}{\bm{B}}

\newcommand{\bD}{\bm{D}}

\newcommand{\bI}{\bm{I}}

\newcommand{\bN}{\bm{N}}

\newcommand{\bS}{\bm{S}}

\newcommand{\bX}{\bm{X}}

\newcommand{\bx}{\bm{x}}

\newcommand{\bz}{\bm{z}}

\newcommand{\bSigma}{\bm{\Sigma}}

\newcommand{\bTheta}{\bm{\Theta}}

\newcommand{\bpi}{\bm{\pi}}
\newcommand{\btheta}{\bm{\theta}}
\newcommand{\bmu}{\bm{\mu}}

\newcommand{\ba}{\bm{a}}

\usepackage{tikz}
\usetikzlibrary{graphs}
\usetikzlibrary{positioning}

\newcommand{\black}{\color{black}}

\RequirePackage{amsthm,amsmath,amsfonts,amssymb}
\RequirePackage[colorlinks,citecolor=blue,urlcolor=blue]{hyperref}
\RequirePackage{graphicx}

\usepackage{cancel}

\makeatletter 

\newcommand*\Neginternal[3]{\mathpalette\Neg@{{#1}{#2}{#3}}}
\newcommand*\Neg@[2]{\Neg@@{#1}#2}
\newcommand*\Neg@@[4]{%
	\mathrel{\ooalign{%
			$\m@th#1#4$\cr
			\hidewidth$\m@th#3{#1}\mkern\muexpr#2*2$\hidewidth\cr
	}}%
}

\newcommand*\negslash[1]{\m@th#1\not\mathrel{\phantom{=}}}
\newcommand*\snegslash[1]{\rotatebox[origin=c]{60}{$\m@th#1-$}}
\newcommand*\ssnegslash[1]{\rotatebox[origin=c]{60}{$\m@th#1{\dabar@}\mkern-7mu{\dabar@}$}}
\newcommand*\sssnegslash[1]{\rotatebox[origin=c]{60}{$\m@th#1\dabar@$}}
\makeatother  

\usepackage{multirow}

\usepackage[ruled,vlined,linesnumbered,lined,boxed,commentsnumbered]{algorithm2e}
\SetKwInput{KwInput}{Input}
\SetKwInput{KwOutput}{Output}

\usepackage{tikz}
\usetikzlibrary{arrows.meta}

\usepackage{soul}	

\theoremstyle{plain}

\usepackage{bm}
\usepackage{bbm}
\usepackage{float}
\usepackage{tikz}
\usepackage{amsthm}
\usepackage{amssymb}
\usepackage{subfig}
\usepackage{longtable}
\usepackage{mathtools,cancel}
\usepackage{url}
\usepackage[ruled,vlined,linesnumbered,lined,boxed,commentsnumbered]{algorithm2e}
\usetikzlibrary{positioning}
\usetikzlibrary{graphs} 
\usetikzlibrary[graphs]
\usepackage{multirow}
\usepackage{lscape}
\usepackage[english]{babel}
\usepackage{amsmath}
\usepackage{authblk}
\usepackage{xr}

\usepackage{tikz}
\usetikzlibrary{positioning}

\RequirePackage[OT1]{fontenc}
\RequirePackage{amsthm,amsmath}

\newcommand{\baa}{\begin{eqnarray}}
	\newcommand{\eaa}{\end{eqnarray}}

\begin{document}

\def\spacingset#1{\renewcommand{\baselinestretch}%
{#1}\small\normalsize} \spacingset{1}


\if0\blind
{
  \title{Joint structure learning and causal effect estimation for categorical graphical models}
  \author[1]{Federico Castelletti \thanks{federico.castelletti@unicatt.it}}
  \author[2]{Guido Consonni \thanks{guido.consonni@unicatt.it}}
  \author[2]{Marco L. Della Vedova \thanks{marco.dellavedova@chalmers.se}}
  \affil[1,2]{Department of Statistical Sciences, Universit\`{a} Cattolica del Sacro Cuore, Milan}
  \affil[3]{Department of Mechanics and Maritime Sciences, Chalmers University of Technology, G\"{o}teborg}
  \maketitle
} \fi

\if1\blind
{
  \bigskip
  \bigskip
  \bigskip
  \begin{center}
    {\LARGE\bf Learning causal effects from categorical graphical models}
\end{center}
  \medskip
} \fi


\begin{abstract}

We consider a
a collection of categorical random variables.
Of special interest is the causal effect on an outcome variable
following  an intervention on another variable.
Conditionally on a Directed Acyclic Graph (DAG), we assume that the joint law of the random variables
can be factorized according to the DAG, where each term is a categorical distribution for the node-variable given a configuration of its parents.
The graph is equipped with  a causal interpretation through the
notion of interventional distribution and the allied \lq \lq do-calculus\rq \rq{}.
From a modeling perspective, the likelihood
is decomposed into a product over nodes and parents of DAG-parameters,
on which a suitably specified collection of Dirichlet priors is assigned.
The overall joint distribution on the ensemble of DAG-parameters is then constructed using  global and local independence.
We account for DAG-model uncertainty and propose a reversible jump Markov Chain Monte Carlo (MCMC)
algorithm which targets the joint posterior over DAGs and DAG-parameters; from the output we are able to recover a full posterior distribution of any causal effect coefficient of interest, possibly summarized by a Bayesian Model Averaging (BMA)
point estimate.
We validate our method through extensive simulation studies, wherein comparisons with alternative state-of-the-art procedures reveal an outperformance in terms of estimation accuracy. Finally, we analyze a dataset relative to a study on depression and anxiety in undergraduate students.

\end{abstract}

\noindent%
{\it Keywords:}
Bayesian inference;
Directed Acyclic Graph,
Categorical data;
Causal inference;
Markov chain Monte Carlo
\vfill

\newpage
\spacingset{1.5} 

\section{Introduction}

Causal inference \citep{Pear:2000, Imbens:Rubin:2015} is a very important area of scientific investigation across a variety of disciplines.
A typical setting  envisages a system of related variables and  addresses the
basic causal question: \lq \lq What is  the effect on a  variable following  an \textit{intervention} 
on another  variable?\rq \rq{}.
A powerful theoretical framework to effectively handle causal inference
is that of a \textit{causal model}, a pair
comprising a graph and a  probability distribution on all the  variables. Specifically the graph is taken to be  directed and acyclic (Directed Acyclic Graph or DAG),   while the  distribution satisfies the Markov factorization of the DAG \citep{Laur:1996, Sadeghi:2017}.
When data are available  the scope of the  causal model is widened to include a  \textit{family} of Markov-probability distributions which are named \textit{observational} because they are meant to describe the joint occurrence of the variables as they naturally arise.
A related formal representation for causal inference is represented by a structural equation model
\citep{Pearl:1995} but is not discussed in the present work.

The term \lq \lq causal\rq \rq{} acquires its meaning when the pair is equipped with the definition  of
\textit{interventional distribution} induced by an external action; the latter is tied to the DAG-factorization and is a cornerstone of the \textit{do calculus} \citep{Pearl:2009}.
A notable feature of the interventional distribution  is that, under suitable assumptions, 
it  can be expressed in terms of the {observational} distribution alone, meaning that causal queries can be  answered  based on observational data, and this is the setting we adopt in this paper. The case in which  both observational and interventional data are available and jointly modeled  is presented in \citet{Hauser:Buehlmann:2015}.

A causal model is predicated on a \textit{given} DAG. In real-world applications however
the generating DAG is unknown and thus needs to be estimated. A difficulty we face is that the true generating DAG is not identifiable from purely observational data because
its conditional independencies can be encoded in
different DAGs which can be grouped into a (Markov) \emph{equivalence class}; identifiability can be  reached but this requires specific distributional assumptions; see for instance \citet{Pete:Buhl:2014}, \citet{Mahdi:Wit:2018}, \citet{Hoyer:et:al:2008}, \citet{Shimizu:et:al:2006} and will not be dealt in this paper.
Because only a Markov equivalence class can be inferred from data, it follows that  there exists a whole
\emph{collection}  of causal effects (one for each DAG in the  class); see
\citet{Maat:etal:2009} for methods to identify these effects in high-dimensional multivariate Gaussian models.

Historically  DAGs were introduced  for probabilistic systems of categorical variables, and in that setting they acquired the name of \textit{Bayesian networks} \citep{Pearl:book:1988:}. The foundations of causality based on the DAG-approach were also mostly developed for discrete/categorical variables \citep{Pear:2000}.
Bayesian causal discovery methods for discrete variables can be traced back to
\citet{Heckerman:1995};  see also \citet{scutari2014bayesian} and \citet{roverato:2017} for a more recent account.
\citet{Madi:Etal:1996} and \citet{Cast:Perl:2004}
and more recently \citet{Castelletti:Peluso:2021}
focus on learning equivalence classes.
A large part of recent methodological research in  causal inference is however framed in the context of continuous multivariate distributions
\citep{Maathuis:Nandy:Review}.

In this paper we develop  a Bayesian method for causal inference when all the variables are categorical  combining  structure learning and inference on causal effects.
We fully account for uncertainty of inference both on the DAG-structure and the main parameters of interest.
Specifically,
Section 2 presents relevant notation, the model formulation and the allied priors; Section 3 specifies the  causal effect as the main parameter of inference;  Section 4 details our computational strategy leading up to a Bayesian Model Averaging estimate of the  target parameter.
The performance of our method,  including comparisons with alternative approaches,  is presented in Section 5, while Section 6 presents an application to depression and anxiety data.
The final section offers a brief discussion together with possible future developments.
Code implementing our methodology is publicly available at \url{https://github.com/FedeCastelletti/bayes\_structure\_causal\_categorical\_graphs}.


\section{Bayesian inference of categorical DAG models}

\subsection{Categorical data and notation}

Let $X = (X_j, j\in V)^\top$, $V=\{1,\dots,q\}$, be a $(q,1)$ vector
%
\black
of categorical random variables with $X_j$ taking values in the corresponding set of levels  $\X_j$, whose generic element (level) is  $x_j$. It follows that $X \in \X \coloneqq \times_{j \in V}\X_j$, the product space generated by the levels of the $q$ variables, whose generic element is $x \in \X$.
For any $x \in \X$, we can consider the joint probability $\pi_{x}=\Pr(X = x \g \bpi)$, where the resulting collection $\bpi=\{\pi_{x},x \in \X\}$ can  be arranged as a \textit{q}-dimensional contingency table of probabilities, where each cell  refers to a specific level $x \in \X$.
For any given $S \subseteq V$ we let $X_S=(X_j, j \in S)$ be  the sub-vector of $X$ with components indexed by $S$, and $x_S\in\X_S\coloneqq \times_{j \in S}\X_j$ one of its levels.
We then let $\pi^S_{x_S}=\Pr(X_S = x_S \g \bpi)$ be the corresponding marginal joint probability for variables in $S$.
We instead write $\theta_{x_j\g x_S}^{\,j\g S}=\Pr(X_j=x_j\g X_S=x_S, \bpi)$ to denote the conditional probability for variable $X_j$ evaluated at $x_j$, given configuration $x_S$ of variables in $S$, $j \notin S$.

Consider now $n$ observations from $X$, $\bx^{(1)},\dots,\bx^{(n)}$, where $\bx^{(i)}=(x_1^{(i)},\dots,x_q^{(i)})^\top$ and $\bx^{(i)}\in\X$, for $i=1,\dots,n$.
For any $x\in\X$, we can compute the count $n_x=\sum_{i=1}^n\mathbbm{1}(\bx^{(i)}=x)$, i.e.~the number of observations that are equal to $x$, and organize the resulting collection of values in a \textit{q}-dimensional contingency table of counts $\bN=\{n_x,x \in \X\}$.
In addition, for any $x_S\in\X_S$, we let $n_{x_S}^S=\sum_{i=1}^n\mathbbm{1}(\bx_S^{(i)}=x_S)$ and $\bN_S=\{n_{x_S}^S,x_S \in \X_S\}$ be the allied $|S|$-dimensional marginal contingency table of counts.

\subsection{Model formulation}

Consider  a Directed Acyclic Graph (DAG) $\D=(V,E)$, with set of nodes $V$, one for each of the $q$ variables, and $E\subseteq V \times V$ its set of directed edges. If $(u,v) \in E$, then $(v,u) \notin E$,  and we say that $\D$ contains the directed edge $u\rightarrow v$,
where $u$  is a \textit{parent} of $v$; equivalently  $v$ is a \textit{child} of $u$.
The set of all parents of $u$ in $\D$ is written $\pa_{\D}(u)$, while  $\fa_{\D}(u)=u\cup \pa_{\D}(u)$ identifies the \textit{family} of $u$.
In the remainder of this section and in Section \ref{sec:causal:effects} we reason \emph{conditionally} on a single given DAG which for simplicity is omitted from our notation.
Under $\D$, and for any level $x\in\X$, the joint probability function of the random vector $X$ factorizes as
\ben
\label{eq:observational:distrib}
\begin{aligned}
	p(x) \,&=\,
	\Pr\big(X_1=x_1,\dots,X_q=x_q) \\
	\,&=\, \prod_{j=1}^q p\big(X_j = x_j\g X_{\pa(j)}=x_{\pa(j)}\big).
\end{aligned}
\een
From a modeling perspective,
given independent realizations $\{\bx^{(i)}, i=1, \ldots,n\}$,
the likelihood function is then
\ben
\label{eq:likelihood}
\begin{aligned}
	p(\bX\g\btheta) \,&=\,
	\prod_{i=1}^n
	\left\{\prod_{x\in \X}
	\left\{\Pr\big(X_1^{(i)}=x_1^{(i)},\dots,X_q^{(i)}=x_q^{(i)} \g \btheta \big)\right\}^{\mathbbm{1}(\bx^{(i)}= \,x)}\right\} \\
	\,&=\, \prod_{j=1}^q\left\{
	\prod_{k\in \X_{\pa(j)}}
	\left\{
	\prod_{m\in \X_j}
	\left\{
	\theta_{m\g k}^{\,j\g \pa(j)}
	\right\}^{n_{(m,k)}^{\fa(j)}}
	\right\}
	\right\},
\end{aligned}
\een
where $\bX$ is the $(n,q)$ observed data matrix whose $i$-th row is $(\bx^{(i)})^\top $.
Notice that Equation \eqref{eq:likelihood} depends on the raw observations $\bX$ through the counts $\bN$ which are the sufficient statistic.

\subsection{Parameter prior distributions}
\label{sec:priors}

We now proceed by assigning a prior distribution to $\btheta$.
Specifically, consider for each $j\in V$ and each $x_{\pa(j)}\in\X_{\pa(j)}$ the allied set of parameters
\be
\left(
\theta_{x_j\g x_{\pa(j)}}^{\,j\g \pa(j)}, x_j \in \X_j
\right)
&\coloneqq& \btheta_{x_{\pa(j)}}^{\,j\g\pa(j)},
\ee
where each element is a  $|\X_j|$-dimensional vector of conditional probabilities for variable $X_j$ given configuration $x_{\pa(j)}$ of its parents.
We introduce the following independence assumptions on the resulting collection of vector-probabilities \citep{Geiger:Heckerman:1997:Dirichlet}:
\begin{itemize}
	\item (G) $\underset{j\in V}\ind \,\btheta_{x_{\pa(j)}}^{\,j\g\pa(j)}$,
	for each parent configuration $x_{\pa(j)}$
	(\textit{global} parameter independence);
	\item  (L) $\underset{x_{\pa(j)} \in \X_{\pa(j)}}\ind \btheta_{x_{\pa(j)}}^{\,j\g\pa(j)}$,
	for each variable $j$
	(\textit{local} parameter independence).
\end{itemize}
Furthermore, we assume for each
$\btheta_{k}^{\,j\g\pa(j)}$, with $j \in V$ and $k \in \X_{\pa(j)}$,
\ben
\btheta_{k}^{\,j\g\pa(j)} \,\sim\,
\textnormal{Dir}\big(\ba_k^{\,j\g\pa(j)}\big),
\een
a Dirichlet distribution with hyperparameter $\ba_k^{\,j\g\pa(j)} =
\big(a_{m\g k}^{\,j\g\pa(j)}>0, m\in \X_j\big)$,
whose probability density function is given by
\ben
\label{eq:prior:dirichlet}
\begin{aligned}
	p\left(\btheta_k^{\,j\g\pa(j)}\right)
	\,&=\,
	\frac
	{\Gamma\left(\sum_{m \in \X_j} a_{m\g k}^{\,j\g\pa(j)}\right)}
	{\prod_{m \in \X_j}\Gamma\left(a_{m\g k}^{\,j\g\pa(j)}\right)}
	\prod_{m \in \X_j}
	\left\{\theta_{m\g k}^{\,j\g\pa(j)}\right\}^{a_{m\g k}^{\,j\g\pa(j)}-1}\\
	\,&=\,
	h\left(\ba_k^{\,j\g\pa(j)}\right)
	\prod_{m \in \X_j}
	\left\{\theta_{m\g k}^{\,j\g\pa(j)}\right\}^{a_{m\g k}^{\,j\g\pa(j)}-1},
\end{aligned}
\een
where $h(\cdot)$ is the prior normalizing constant.
The resulting collection of Dirichlet distributions, together with (G) and (L),  determines a prior on the overall DAG-parameter
\black
\ben
\label{eq:overall-DAG-parameter}
\btheta \,=\, \left\{\btheta_{k}^{\,j\g\pa(j)}, \, j \in V, \, k \in \X_{\pa(j)}\right\}
\een
which
factorizes as
\ben
\begin{aligned}
	\label{eq:prior:theta}
	p\left(\btheta\right)
	\,&=\,
	\prod_{j \in V}
	\left\{
	\prod_{k \in \X_{\pa(j)}}
	p\left(\btheta_k^{\,j\g\pa(j)}\right)
	\right\} \nonumber \\
	\,&=\,
	\prod_{j \in V}
	\left\{
	\prod_{k \in \X_{\pa(j)}}
	\textnormal{pDir}\left(\btheta_k^{\,j\g\pa(j)} \gbig \ba_k^{\,j\g\pa(j)}\right)
	\right\}.
\end{aligned}
\een
The choice of the hyperparameters in \eqref{eq:prior:theta} 
requires care especially
when several DAGs are entertained and the purpose is DAG model selection.
In particular, because observational data cannot distinguish between Markov equivalent DAGs, the prior on the  parameter $\btheta$ should guarantee that
any two equivalent DAGs are assigned the same \emph{marginal} likelihood; this is the rationale behind the procedure for prior elicitation introduced by \citet{Heckerman:1995} leading to their Bayesian Dirichlet Equivalent uniform score (BDEu);
see also \citet{Geig:Heck:2002}.
Specifically, these authors show that the default choice
\ben
\label{eq:dirichlet:hyperparameter}
a_{m\g k}^{\,j\g\pa(j)} = \frac{a}{|\X_{\fa(j)}|}, \quad
j\in V, \quad m \in \X_j, \quad k \in \X_{\pa(j)},
\een
with $a>0$, guarantees DAG score equivalence.

\section{Causal effects}
\label{sec:causal:effects}

The DAG factorization \eqref{eq:observational:distrib} is also called the \textit{observational} (or \textit{pre-intervention}) distribution.
Consider now two variables, $X_v$ and $X_h \coloneqq Y$ $(h\ne v)$ where the latter is a response of interest.
We are interested in the (total) \textit{causal effect} on $Y$ of an intervention on $X_v$. 
In particular we consider a \emph{hard} intervention on $X_v$, consisting in the action of forcing its value to a given level $\tilde{x}$,   denoted  $\textnormal{do}(X_v=\tilde{x})$.
Under a hard intervention, the \textit{post-intervention} distribution \citep{Pear:2000} is obtained through the truncated factorization
\ben
\label{eq:post:intervention}
p\big(x\g\textnormal{do}(X_v = \tilde{x})\big) =
\begin{cases}
	\prod\limits_{j \ne v} p\big(X_j = x_j\g X_{\pa(j)}=x_{\pa(j)}\big)
	& \text{if }  x_v=\tilde{x}
	\\
	\,\, 0 & \text{otherwise},
\end{cases}
\een
where each term $p(X_j = x_j\g\cdot)$ is the corresponding (pre-intervention) conditional distribution of Equation \eqref{eq:observational:distrib}.
Assuming for simplicity that both $X_v$ and $Y$ are binary taking values in $\{0,1\}$, the causal effect on $Y$ resulting from an intervention on $X_v$ can be defined as
\ben
\label{eq:causal:effect}
c_v =
\vat\big(Y\g \textnormal{do}(X_v=1)\big)-
\vat\big(Y\g \textnormal{do}(X_v=0)\big).
\een
Moreover, it can be shown \citep[Theorem 3.2.3]{Pear:2000} that
\ben
\begin{aligned}
	\label{eq:cv}
	c_v =
	\sum_{k \in \X_{\pa(v)}} \vat\big(Y\g & X_v=1, X_{\pa(v)}=k\big)
	\Pr\big(X_{\pa(v)}=k\big)\\
	-&
	\sum_{k \in \X_{\pa(v)}} \vat\big(Y\g X_v=0, X_{\pa(v)}=k\big)
	\Pr\big(X_{\pa(v)}=k\big),
\end{aligned}
\een
where the expectation can be alternatively written in terms of probabilities because of the binary nature of $Y$. Equation \eqref{eq:cv} uses the set of parents as an adjustment set; however alternative sets are also available \citep{Pearl:2009,Henckel:et:al:2022}.
From a modeling perspective,
the causal effect
can be written as
\ben
\label{eq:gamma:v}
\gamma_v(\btheta) \,=
\sum_{k \in \X_{\pa(v)}}
\left\{
\left(
\theta_{1\g (1,k)}^{\,Y\g \fa(v)} -
\theta_{1\g (0,k)}^{\,Y\g \fa(v)}
\right)
\theta_{k}^{\,\pa(v)}
\right\}.
\een
Notice that the $\theta$-parameters involved in
\eqref{eq:gamma:v} are not the components
of the overall DAG-parameter $\btheta$ in \eqref{eq:overall-DAG-parameter} because the conditional distribution of $Y \g X_{\fa(v)}$ does not appear in general in the factorization \eqref{eq:likelihood}.
Yet
$\gamma_v$ is a function of $\btheta$, so that inference on $\gamma_v$ can be retrieved from the posterior distribution  of $\btheta$, which is the subject of the next section.
When $X_v$ is polytomous one  can define a battery of causal effects. Typically one would choose  a \textit{reference} level for $X_v$, $\tilde {m}$ say, and then apply \eqref{eq:causal:effect} for pairs $(X_{v}=m, X_v=\widetilde {m})$ with $m \neq \widetilde {m}$.
On the other hand when the levels of the response  $Y$ are more than two, the conditional expectation in \eqref{eq:causal:effect} should be replaced  by the  probability that $Y$ attains a suitable  benchmark level.
Alternatively,  a collection of $Y$-level dependent causal effects can be computed and then analyzed to gauge sensitivity.

\section{Posterior inference}

Let $\mathcal{S}_q$ be the set of all DAGs with $q$ nodes. In this section we also regard DAG $\D$ as uncertain and introduce a Markov Chain Monte Carlo (MCMC) scheme for posterior inference on $(\D,\btheta)$.
Let $p(\D)$ be a prior on $\D \in \mathcal{S}_q$ that we specify in Section \ref{sec:prior:DAG}.
Our target is the joint posterior distribution
\ben
\label{eq:posterior:dag:theta}
p(\btheta,\D\g\bX)\propto
p(\bX\g\btheta,\D)\,p(\btheta\g \D)\,p(\D),
\een
where we now emphasize the dependence on DAG $\D$ both in the likelihood and in the prior on $\btheta$.

\subsection{Prior on DAG $\D$}
\label{sec:prior:DAG}

We assign a prior on DAGs belonging to $\mathcal{S}_q$ through a Beta-Binomial distribution on the number of edges in the graph.
Specifically, for a given DAG $\D=(V,E)\in\mathcal{S}_q$, let $\bS^{\D}$ be the $0-1$ \emph{adjacency matrix} of its skeleton, that is the underlying undirected graph obtained after removing the orientation of all its edges. For each $(u,v)$-element of $\bS^{\D}$, we have $\bS_{u,v}^{\D}=1$ if and only if $(u,v)\in E$ or $(v,u)\in E$, zero otherwise.
Conditionally on a prior probability of inclusion $\eta\in(0,1)$ we assume, for each $u>v$,
$
\bS_{u,v}^{\D} \g \eta \overset{\textnormal{iid}}\sim \text{Ber}(\eta),
$
which implies
\ben
p(\bS^{\D}\g \eta)=\eta^{|\bS^{\D}|} (1-\eta)^{\frac{q(q-1)}{2}-|\bS^{\D}|},
\een
where $|\bS^{\D}|$ is the number of edges in $\D$ (equivalently in its skeleton) and $q(q-1)/2$ is the maximum number of edges in a DAG on $q$ nodes.
We then assume $\eta\sim \textnormal{Beta}(c,d)$, so that, by integrating out $\eta$, the resulting prior on
$\bS^{\D}$ is
\black
\ben
\label{eq:prior:multiplicity}
p(\bS^{\D}) = \frac
{\Gamma \left(|\bS^{\D}| + c\right)\Gamma \left(\frac{q(q-1)}{2} - |\bS^{\D}| + d \right)}
{\Gamma \left(\frac{q(q-1)}{2} + c + d\right)}
\cdot
\frac
{\Gamma \left(c + d\right)}
{\Gamma \left(c\right)\Gamma \left(d\right)}.
\een
Finally, we set
$
p(\D)\propto p(\bS^{\D})$ for each $\D\in\mathcal{S}_q
$.

\subsection{MCMC scheme}

Our sampler is based on a reversible jump MCMC algorithm which takes into account the Partial Analytic Structure (PAS; \citet{Godsill:2012}) of the prior on $\btheta$ (Section \ref{sec:priors}). Specifically, it implements two steps which iteratively update $\D$ and $\btheta$ by sampling from their full conditional distributions.

\subsubsection{Update of $\D$}

To sample from the full conditional distribution of DAG $\D$ we adopt a Metropolis Hastings (MH) scheme. This requires the construction of a suitable proposal distribution which determines the transitions between graphs in $\mathcal{S}_q$, the set of all DAGs on $q$ nodes.
Given a DAG $\D$ we consider three types of operators which locally modify $\D$ by $(i)$ inserting a directed edge $u\rightarrow v$ (Insert $u\rightarrow v$),
$(ii)$ deleting a directed edge $u\rightarrow v$ (Delete $u\rightarrow v$),
$(iii)$ reversing of a directed edge $u\rightarrow v$ (Reverse $u\rightarrow v$).
An operator is \textit{valid} if the resulting graph is a DAG.
Let $\mathcal{O}_{\D}$ be the set of all valid operators on $\D$, $|\mathcal{O}_{\D}|$ its size. A new DAG $\widetilde\D$ is obtained by uniformly drawing an element from $\mathcal{O}_{\D}$ and applying it to $\D$; the proposal distribution determining a transition from $\D$ to $\widetilde\D$ is then $q(\widetilde\D\g\D)=1/|\mathcal{O}_{\D}|$.

Notice that $\widetilde\D$ only differs \textit{locally} from $\D$, because it is obtained by inserting, deleting or reversing a single edge $u \rightarrow v$.
Accordingly, consider two DAGs $\D=(V,E)$, $\widetilde\D=(V,\widetilde E)$ such that $\widetilde E=E\setminus\{(u,v)\}$
and let
$\btheta$, $\widetilde\btheta$ be the corresponding DAG-dependent parameters.
Because of the structure of our prior (Section \ref{sec:priors}) the two sets of parameters differ with regard to their $v$-th component only, namely
$\left\{\btheta_k^{\,v\g\pa(v)}, k \in \X_{\pa_{\D}(v)}\right\}=\btheta_v$ and $\left\{\widetilde\btheta_k^{\,v\g\pa(v)}, k \in \X_{\pa_{\widetilde\D}(v)}\right\}=\widetilde\btheta_v$ respectively.
Let also $\btheta_{-v}=\btheta\setminus\btheta_v$; similarly for $\widetilde \btheta_{-v}$.
Moreover, because $\pa_{\D}(j)=\pa_{\widetilde\D}(j)$ for all $j\ne v$, the remaining sets of parameters are componentwise equivalent.
The acceptance probability for $\widetilde\D$ under a PAS algorithm
is then
$\alpha_{\widetilde\D}=\min\{1;r_{\widetilde\D}\}$,
where
\ben
\label{eq:ratio:PAS:DAG}
r_{\widetilde\D}
&=&\frac{p(\widetilde\D\g\widetilde\btheta_{-v},\bX)}
{p(\D\g\btheta_{-v},\bX)}
\cdot
\frac{q(\D\g\widetilde\D)}{q(\widetilde\D\g\D)}
\nonumber \\
&=&
\frac{p(\bX,\widetilde\btheta_{-v}\g \widetilde\D)}
{p(\bX,\btheta_{-v}\g \D)}
\cdot\frac{p(\widetilde\D)}{p(\D)}
\cdot\frac{q(\D\g\widetilde\D)}{q(\widetilde\D\g\D)}.
\een
Therefore, we require to compute for DAG $\D$
\be
p(\bX,\btheta_{-v}\g \D)
&=&
\int_{\bTheta_v}
p(\bX\g\btheta,\D)p(\btheta\g\D) \, d \btheta_v,
\ee
and similarly for $\widetilde\D$,
where importantly $p(\bX\g\btheta,\D)$ and $p(\btheta\g\D)$ admit the factorizations in
\eqref{eq:likelihood} and \eqref{eq:prior:theta} respectively.
Accordingly, we can write
\ben
\label{eq:PAS:joint:DAG}
p(\bX,\btheta_{-v}\g \D)
&=&
\prod_{j\ne v}
\left\{
\prod_{k\in \X_{\pa(j)}}
\left\{
h\left(\ba_k^{\,j\g\pa(j)}\right)
\prod_{m\in \X_j}
\left\{
\theta_{m\g k}^{\,j\g \pa(j)}
\right\}^{a_{m\g k}^{\,j\g\pa(j)}+n_{(m,k)}^{\fa(j)}-1}
\right\}
\right\} \nonumber
\\
&\cdot&
\int_{\bTheta_v}
\left\{
\prod_{k\in \X_{\pa(v)}}
\left\{
h\left(\ba_k^{\,v\g\pa(v)}\right)
\prod_{m\in \X_v}
\left\{
\theta_{m\g k}^{\,v\g \pa(v)}
\right\}^{a_{m\g k}^{\,v\g\pa(v)}+n_{(m,k)}^{\fa(v)}-1}
\right\}
\right\}d \btheta_v \nonumber
\een
Now, letting $g\big(\bX,\btheta_{-v}\big)$ be the first term of the previous expression, and using Equation \eqref{eq:prior:dirichlet}, we can write
\ben
\begin{aligned}
	p(\bX,\btheta_{-v}\g \D)
	\,&=\,
	g\big(\bX,\btheta_{-v}\big)
	\cdot
	\prod_{k\in \X_{\pa(v)}}
	\frac
	{h\left(\ba_k^{\,v\g\pa(v)}\right)}
	{h\left(\ba_k^{\,v\g\pa(v)}+\bN_{\fa(v)}^k\right)} \nonumber
	\\
	\,&=\,
	g\big(\bX,\btheta_{-v}\big) \cdot
	m\big(\bX_v\g\bX_{\pa(v)},\D\big),
\end{aligned}
\een
where $\bN_{\fa(v)}^k$ is the contingency table of counts for variables in $\fa(v)$, obtained by including only those observations corresponding to configuration $k \in \X_{\pa(v)}$.
Therefore, the acceptance ratio \eqref{eq:ratio:PAS:DAG} simplifies to
\ben
\label{eq:ratio:PAS:DAG:simplified}
r_{\widetilde\D} =
\frac
{m\big(\bX_v\g\bX_{\widetilde\pa(v)},\widetilde\D\big)}
{m\big(\bX_v\g\bX_{\pa(v)},\D\big)}
\cdot\frac{p\big(\widetilde\D\big)}{p\big(\D\big)}
\cdot\frac{q\big(\D\g\widetilde\D\big)}{q\big(\widetilde\D\g\D\big)},
\een
with $\pa(v) \coloneqq \pa_{\D}(v), \widetilde\pa(v) \coloneqq \pa_{\widetilde \D}(v)$ and all terms available in closed-form expression.

\subsubsection{Update of \boldmath{$\theta$}}

Conditionally on the updated DAG $\D$,
in the second step of the PAS algorithm we sample the DAG-dependent parameter $\btheta$ from its full conditional distribution
\ben
\begin{aligned}
	p(\btheta\g\D,\bX)
	\,&\propto\,
	p(\bX\g\btheta,\D)\,p(\btheta\g \D) \\
	\,&\propto\,
	\prod_{j \in V}
	\left\{
	\prod_{k \in \X_{\pa(j)}}
	\left\{
	\prod_{m \in \X_j}
	\left\{\theta_{m\g k}^{\,j\g\pa(j)}\right\}^{a_{m\g k}^{\,j\g\pa(j)}+n_{(m,k)}^{\fa(v)}-1}
	\right\}
	\right\} \\
	\,&=\,
	\prod_{j \in V}
	\left\{
	\prod_{k \in \X_{\pa(j)}}
	\textnormal{pDir}\left(\btheta_k^{\,j\g\pa(j)} \gbig \ba_k^{\,j\g\pa(j)}+\bN_{\fa(j)}^k\right)
	\right\},
\end{aligned}
\een
corresponding to a product of independent (posterior) Dirichlet distributions. Direct sampling from the full conditional of $\btheta$ is therefore straightforward.

\subsection{Posterior summaries}

Output of our MCMC scheme is a collection of DAGs and DAG parameters
$\big\{\big(\btheta^{(1)},\D^{(1)}\big),$ $\dots,\big(\btheta^{(S)},\D^{(S)}\big)\big\}$,
approximately sampled from the posterior \eqref{eq:posterior:dag:theta}, where $S$ the number of final MCMC iterations.
An approximate marginal posterior distribution over the DAG space $\mathcal{S}_q$ can be computed as
\ben
\label{eq:posterior:dag}
\widehat{p}(\D\g\bX)=\frac{1}{S}\sum_{s=1}^S\mathbbm{1}\big(\D^{(s)}=\D\big)
\een
for any $\D \in \mathcal{S}_q$, where $\mathbbm{1}(\cdot)$ is the indicator function, and whose expression corresponds to the MCMC frequency of visits of $\D$.
In addition, for any directed edge $(u, v)$, we can estimate a marginal Posterior Probability of edge Inclusion (PPI) as
\ben
\label{eq:PPI}
\widehat{p}(u\rightarrow v\g\bX)=\frac{1}{S}\sum_{s=1}^S\mathbbm{1}\big(u\rightarrow v \in \D^{(s)}\big),
\een
where $\mathbbm{1}\big(u\rightarrow v \in \D^{(s)}\big)=1$ if $\D^{(s)}$ contains $u \rightarrow v$, $0$ otherwise.
Starting from the previous quantities, single DAG estimates summarizing the MCMC output can be recovered: a Maximum A Posteriori (MAP) estimate, corresponding to the DAG with the highest posterior probability \eqref{eq:posterior:dag} or a Median Probability Model (MPM) estimate, obtained by including only those edges whose PPI \eqref{eq:PPI} is greater that $0.5$.

For a given node $s\in\{2,\dots,q\}$, consider now the causal effect of $\do(X_s = \tilde{s})$ on $Y$,
represented by the parameter
$\gamma_v(\btheta)$ in \eqref{eq:gamma:v}.
For each draw $\btheta^{(s)}$ from the posterior \eqref{eq:posterior:dag:theta}, we can first recover $\gamma_v\big(\btheta^{(s)}\big)$ using Equation \eqref{eq:gamma:v}.
An estimate of $\gamma_v(\btheta)$ is then
\ben
\label{eq:gamma:BMA}
\widehat\gamma_v^{BMA}=\frac{1}{S}\sum_{s=1}^S\gamma_v\big(\btheta^{(s)}\big),
\een
which implicitly performs Bayesian Model Averaging (BMA)
through the MCMC frequencies of the visited DAGs.

\section{Simulation study}


\subsection{Scenarios}

We first illustrate the performance of our methodology through simulation.
Specifically, we consider different scenarios in which we vary the number of variables $q\in\{10,20\}$ and the sample size $n \in \{100,200,500,1000\}$.
For each choice of $q$ we  randomly generate $G=50$ DAGs with probability of edge inclusion $2/q$.
Categorical datasets of size $n$ are then generated as follows.
Each DAG $\D$ defines a data generating process which in a Gaussian setting we can write as
\ben
\label{eq:generating:process}
Z_j^{(i)} = \sum_{u\in \pa_\D(j)} \beta_{j,u}Z_u^{(i)} + \varepsilon_j^{(i)},
\een
for $i=1,\dots,n$ and $v=1,\dots,q$, where $\varepsilon_j^{(i)}\sim \N(0,\sigma_j^2)$ independently.
For each $j$ we fix $\sigma^2_j=1$, while regression coefficients $\beta_{j,u}$ are uniformly chosen in the interval $[-1,-0.1] \cup [0.1,1]$; see also \citet{Pete:Buhl:2014}.
Following \eqref{eq:generating:process}, the joint distribution of $(Z_1,\dots,Z_q)$ is then $\N_q(\bzero,\bSigma)$, with $\bSigma=(\bI_q - \bB)^{-\top}\bD(\bI_q - \bB)^{-1}$, where $\bI_q$ is the $(q,q)$ identity matrix, $\bB$ is the $(q,q)$ matrix with $(j,u)$-element equal to $\beta_{j,u}$, while $\bD=\diag(\sigma_1^2,\dots,\sigma_q^2)$.
Next we generate  $n$ multivariate Gaussian observations from \eqref{eq:generating:process}; a categorical dataset consisting of $n$ observations from $q$ binary variables is then obtained by discretization as
\begin{equation}
	\label{eq:discretize:gaussian}
	X_{j}^{(i)}=
	\begin{cases}
		\,\, 1 & \text{if } Z_j^{(i)}\geq 0, \\
		\,\, 0 & \text{if } Z_j^{(i)} < 0,
	\end{cases}
\end{equation}
where $X_{j}^{(i)}$ is the random variable whose realization is  the $(i,j)$-entry of the $(n,q)$ categorical data matrix $\bX$.
The causal effect of $\textnormal{do}(X_v=\tilde{x})$ on $Y$,  as defined in Equation \eqref{eq:causal:effect},  can be written as
\ben
\label{eq:causal:effect:bis}
\begin{aligned}
	c_v =
	\sum_{k \in \X_{\pa(v)}} \Big\{
	\Big[\Pr&\big(Y=1\g X_v=1,X_{\pa(v)}=k\big) \\
	-& \Pr\big(Y=1\g X_v=0,X_{\pa(v)}=k\big)
	\Big]
	\cdot \Pr\big(X_{\pa(v)}=k\big)\Big\}.
\end{aligned}
\een
Consider first the conditional probability $\Pr\big(Y=1\g X_v=1,X_{\pa(v)}=k\big)$ in
\eqref{eq:causal:effect:bis}.
This can be written as
\ben
\label{eq:conditional:marginal}
\Pr\big(Y=1\g X_v=1,X_{\pa(v)}=k\big)
=
\frac
{\Pr\big(Y=1, X_v=1,X_{\pa(v)}=k\big)}
{\Pr\big(X_v=1,X_{\pa(v)}=k\big)}.
\een
Let now $h_{-1}=-\infty,h_0=0,h_1=\infty$ and for given $k=(k_j)_{j \in \pa(v)}$ such that each $k_j \in \{0,1\}$, let $\mathcal{I}^k=\times_{k_j \in k}(h_{k_j-1},h_{k_j})$.
Because of \eqref{eq:generating:process} and \eqref{eq:discretize:gaussian}, the two joint probabilities in \eqref{eq:conditional:marginal} can be written as
\ben
\label{eq:conditional:marginal:bis}
	\Pr\big(Y=1, X_v=1,X_{\pa(v)}=k\big)
	&=&
	\int_0^\infty
	\int_0^\infty
	\int _{\mathcal{I}^k}
	\phi\big(\bz_{(y,\fa(v))}\g \bzero,\bSigma_{(y,\fa(v)),(y,\fa(v))}\big)\, d\bz_{\pa(v)} \,dz_v \,dz_y, \nonumber \\
	\Pr\big(X_v=1,X_{\pa(v)}=k\big)
	&=&
	\int_0^\infty
	\int _{\mathcal{I}^k}
	\phi\big(\bz_{\fa(v)}\g \bzero,\bSigma_{\fa(v),\fa(v)}\big)\, d\bz_{\pa(v)} \,dz_v,
\een
where $\phi(\bmu, \bSigma)$ denotes the p.d.f.~of a multivariate normal distribution with mean $\bmu$ and covariance matrix $\bSigma$.
To compute the
conditional probability $\Pr\big(Y=1\g X_v=0,\bX_{\pa(v)}=k\big)$ in
\eqref{eq:causal:effect:bis}
simply change the limits of the integral w.r.t.~$z_v$
in \eqref{eq:conditional:marginal:bis} to $(-\infty,0)$.
The true causal effect $c_v$ in \eqref{eq:causal:effect:bis} is then computed for each node $v\in\{2,\dots,q\}$ with node  $Y=X_1$ as the response.

\subsection{Results}

We apply our MCMC scheme to approximate the joint posterior distribution in \eqref{eq:posterior:dag:theta}.
To this end, we let the  number of MCMC iterations $S$ vary in the set $ \{5000,10000\}$ for respectively $q \in\{10,20\}$, disregarding from the output a burn-in period of size $B \in \{1000,2000\}$ for the two values of $q$ respectively.
Moreover, we set the common hyperparameter of the Dirichlet prior in \eqref{eq:dirichlet:hyperparameter} as $a=1$ and $c=d=1$ in the $\textnormal{Beta}(c,d)$ prior for the probability of edge inclusion $\eta$ leading to the prior on DAG-space $p(\D)$; see Section \ref{sec:prior:DAG}.

We start by evaluating the global performance of our method in learning the underlying graphical structure.
Specifically, we first estimate the posterior probabilities of edge inclusion as in \eqref{eq:PPI} for each pair of distinct nodes $(u,v)$ and produce an MPM estimate of the DAG, $\widehat\D$.
The latter is compared with the true DAG $\D$ in terms of sensitivity (SEN) and specificity (SPE)
indexes, respectively defined as
\be
SEN = \frac{TP}{TP + FN}, \quad SPE = \frac{TN}{TN + FP},
\ee
where $TP, TN, FP, FN$ are the numbers of true positives, true negatives, false positives and false negatives, which can be recovered from the 0-1 adjacency matrix of the estimated graphs.
As an overall summary, we also consider the Structural Hamming Distance (SHD), defined as the number of insertions, deletions of flips needed to transform the estimated graph into the true graph.
Results, averaged w.r.t.~the $G=50$ simulations under each scenario defined by $q$ and $n$, are summarized in Table \ref{tab:measures:structure}.
Both the SHD and SEN metrics suggest that the accuracy of our method in recovering the true DAG improves as the number of available data grows; moreover, the SPE index attains high levels even for the smallest value of $n$,  and  is essentially stable as \black the sample size grows; accordingly, the method shows an overall appreciable performance.

\begin{table}
	\centering
	\begin{tabular}{cccccc}
		\hline
		& & $n = 200$ & $n = 500$ & $n = 1000$ & $n = 2000$ \\
		\hline
		\multirow{3}{1.5cm}{$q = 10$} & SHD & 6.35 & 5.22 & 4.50 & 4.15 \\
		& SEN & 56.15 & 69.62 & 78.28 & 82.36 \\
		& SPE & 96.23 & 95.92 & 95.47 & 95.70 \\
		\hline
		\multirow{3}{1.5cm}{$q = 20$} & SHD & 15.47 & 12.55 & 12.10 & 11.40 \\
		& SEN & 50.27 & 66.15 & 73.02 & 74.37 \\
		& SPE & 98.10 & 97.95 & 97.50 & 97.61 \\
		\hline
	\end{tabular}
	\vspace{0.2cm}
	\caption{\label{tab:measures:structure} \small Simulations. Average (w.r.t.~50 simulations) Structural Hamming Distance (SHD), Sensitivity (SEN) and Specificity (SPE) indexes, computed under each scenario defined by number of variables $q \in \{10,20\}$ and sample size $n\in\{200,500,1000,2000\}$.}
\end{table}

We now consider causal effect estimation. To this end, we produce the collection of BMA estimates $\widehat\gamma_v^{BMA}$, $v \in \{2,\dots,q\}$ according to Equation \eqref{eq:gamma:BMA}.
Let now $c_v$ be the true causal effect;
Next we  compare each BMA estimate with the corresponding true  causal effect
$c_v$
and compute the Absolute Error (AE)
\ben
\label{eq:MAE:causal}
AE_v = |c_v - \widehat\gamma_v^{BMA}|.
\een
Results are summarized in Table \ref{tab:measures:causal}, where we report for each value of $q$ and $n$ the average value of AE $\times$ 100 (computed across the $50$ simulated DAGs and nodes $v = 2,\dots,q$).
By increasing the sample size the difference between estimated and true causal effect progressively reduces.
In particular, the average absolute error, 
is  around $0.02$ in the $n=200$ scenario when $q=20$.
This quantity is at most 2\% relative to the maximum potential range of $c_v$.


\begin{table}
	\centering
	\begin{tabular}{ccccc}
		\hline
		& $n = 200$ & $n = 500$ & $n = 1000$ & $n = 2000$ \\
		\hline
		\multirow{1}{1.5cm}{$q = 10$} & 4.46 & 3.70 & 3.50 & 3.28 \\
		\hline
		\multirow{1}{1.5cm}{$q = 20$} & 2.17 & 1.80 & 1.74 & 1.65 \\
		\hline
	\end{tabular}
	\vspace{0.2cm}
	\caption{\label{tab:measures:causal} \small Simulations. Average (w.r.t.~50 simulations and intervened nodes) Absolute Error (AE) between true and estimated causal effect (values multiplied by 100), computed under each scenario defined by number of variables $q \in \{10,20\}$ and sample size $n\in\{200,500,1000,2000\}$.}
\end{table}

\subsection{Comparison with the PC algorithm and IDA approach}

In this section we compare the performance of our Bayesian methodology with the IDA (Identification when DAG is Absent) approach of \citet{Maat:etal:2009}, originally introduced for Gaussian data and adapted to a categorical setting in \citet{Kalish:et:al:2010}.
IDA estimates first a Completed Partially Directed Acyclic Graph (CPDAG) using the PC algorithm \citep{Spir:Glym:Sche:2000,Kalish:Buehlmann:2007}.
The latter is based on a sequence of conditional independence tests that we implement for significance level $\alpha\in\{1\%,5\%,10\%\}$.
The resulting CPDAG represents a Markov equivalence class of DAGs; although these are equivalent in terms of conditional independencies, they can lead in principle to distinct causal effects for the same intervention.
Accordingly, \citet{Maat:etal:2009} propose two different strategies for causal effect estimation.
The first enumerates all DAGs in the equivalence class and for each one estimates the causal effect. As this approach is computationally expensive, even for moderate values of $q$, a second algorithm (hereinafter considered),  which only  outputs  the \emph{distinct} causal effects within a given equivalence class,  is implemented.
Finally, an average causal effect, computed across all distinct causal effects compatible with the estimated CPDAG, is returned.
Each of the distinct causal effect coefficients is computed as in Equation \eqref{eq:gamma:v} upon replacing marginal and conditional probability with the corresponding sample proportions.
We refer to the resulting estimate as $\gamma_v^{IDA}$.
Finally, notice that the PC algorithm provides a CPDAG estimate, rather than a DAG. For comparison purposes we then recover from our MPM DAG estimate the representative CPDAG.
\black

\begin{figure}
	\begin{center}
		\begin{tabular}{cc}
			$q = 10 \quad\quad $ & $q = 20 \quad\quad $ \\
			\includegraphics[scale=0.62]{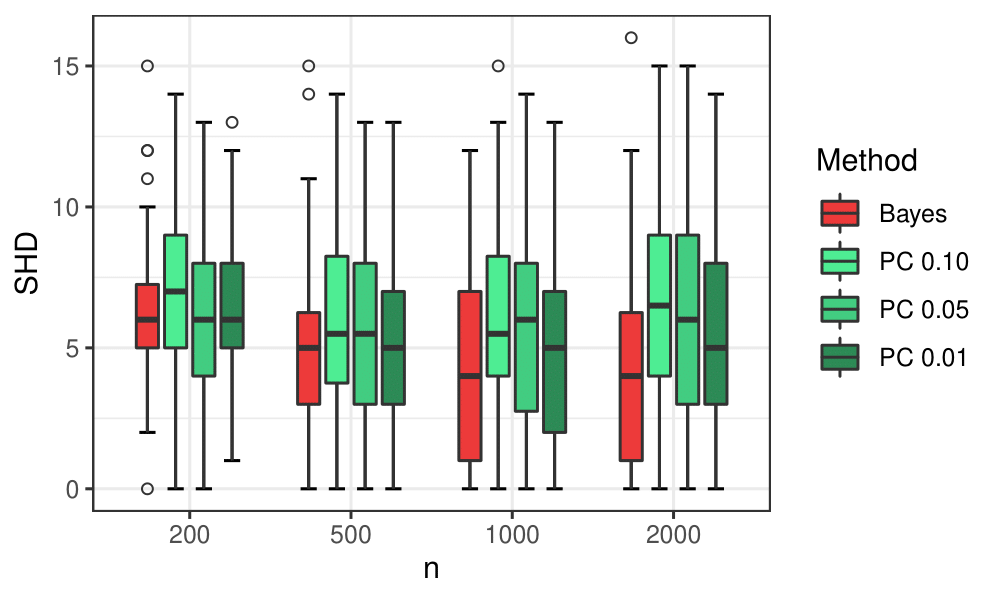} &
			\includegraphics[scale=0.62]{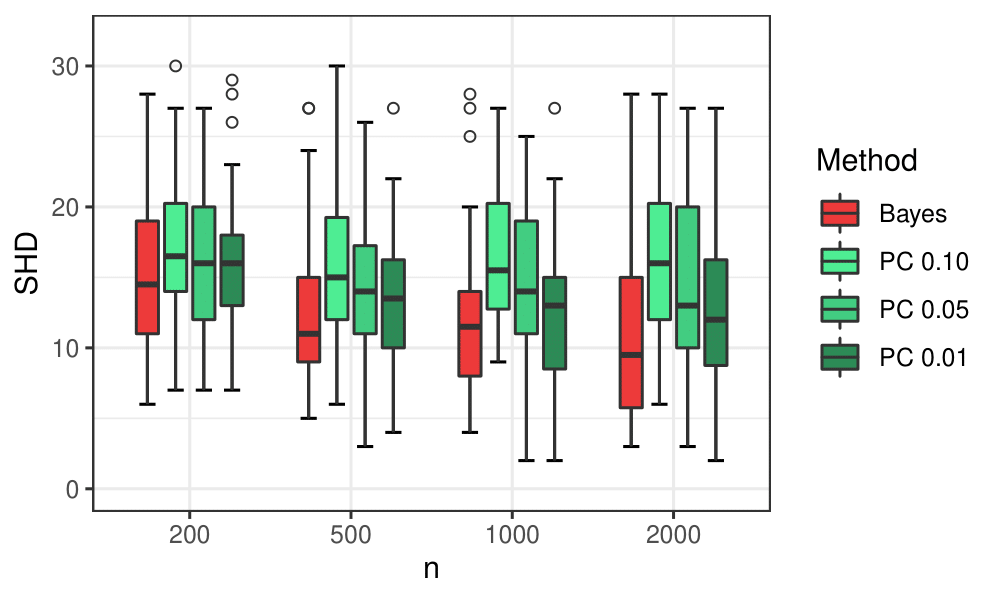}
		\end{tabular}
		\caption{\small Simulations. Structural Hamming Distance (SHD) between true and estimated CPDAGs for number of nodes $q\in\{10,20\}$ and increasing samples sizes $n\in\{200,500,1000,2000\}$. Methods under comparison are: our Bayesian proposal (Bayes) leading to the MPM CPDAG estimate, and the PC algorithm implemented for significance levels $\alpha\in\{0.10,0.05,0.01\}$ (respectively PC 0.10, PC 0.05, PC 0.01).}
		\label{fig:shds}
	\end{center}
\end{figure}

Figure \ref{fig:shds} summarizes the distribution of SHD computed across the $50$ simulations under each method and for different values of $q$ and $n$.
In general, it appears that all methods improve their performance as the sample size grows, with the exception of the PC algorithm which slightly worsens when $n$ increases from $1000$ to $2000$. We remark that our Bayesian method which outputs an MPM-based CPDAG is highly competitive with all three versions of PC and shows an overall better performance across sample sizes when considering the median value of the distribution, while variability is comparable to mildly larger.
\black

\begin{figure}
	\begin{center}
		\begin{tabular}{cc}
			$q = 10 \quad\quad $ & $q = 20 \quad\quad $ \\
			\includegraphics[scale=0.62]{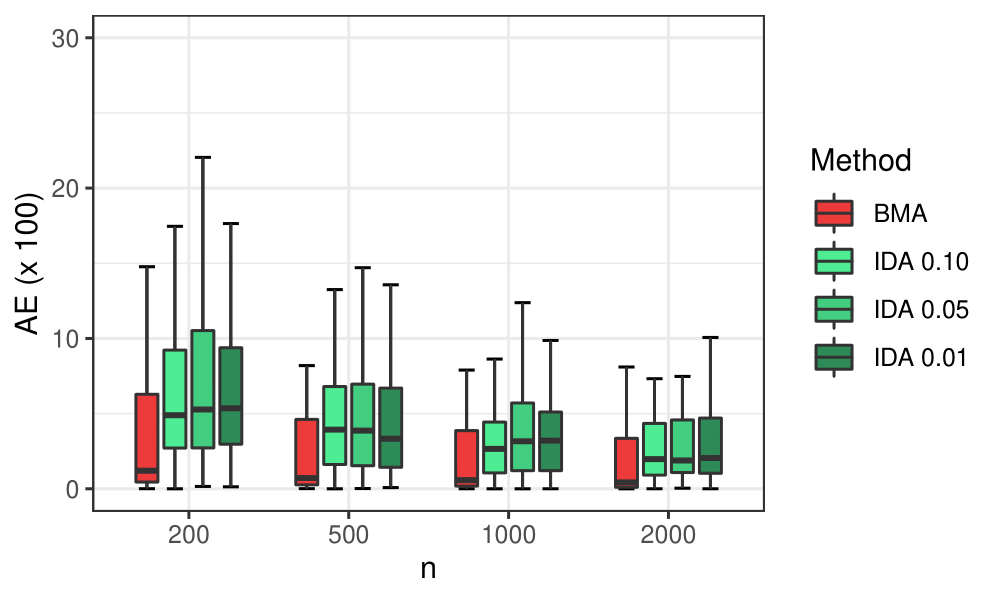} &
			\includegraphics[scale=0.62]{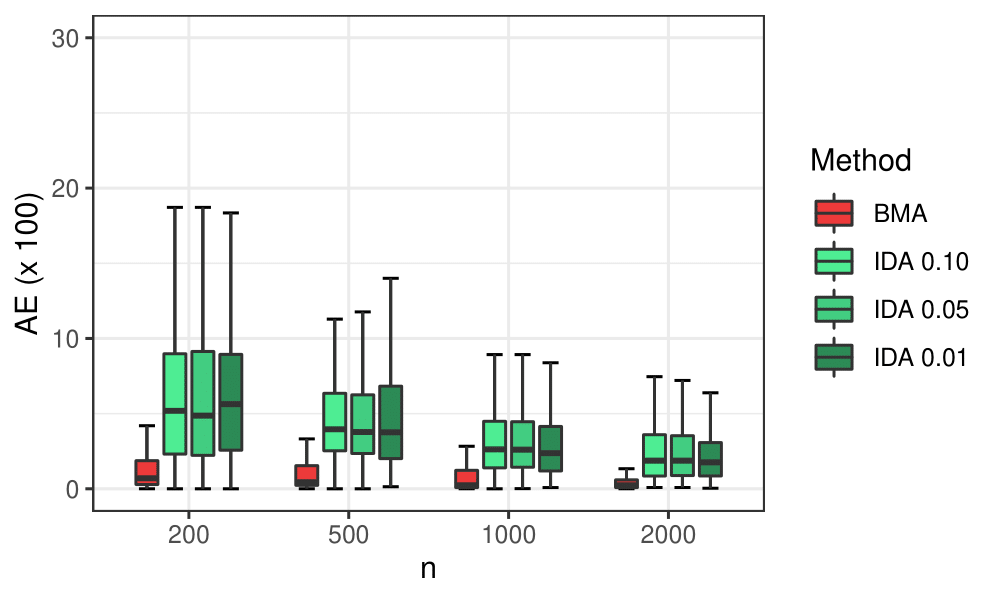}
		\end{tabular}
		\caption{\small Simulations. Absolute Error (AE) between true and estimated causal effects (values multiplied by 100) for number of nodes $q\in\{10,20\}$ and increasing samples sizes $n\in\{200,500,1000,2000\}$. Methods under comparison are: our Bayesian proposal with the BMA causal effect estimate (BMA), and the IDA method based on the PC algorithm implemented for significance levels $\alpha\in\{0.10,0.05,0.01\}$ (respectively IDA 0.10, IDA 0.05, IDA 0.01).}
		\label{fig:ae}
	\end{center}
\end{figure}

Finally, we consider causal effect estimation and report
in Figure \ref{fig:ae} the distribution of the Absolute Error (AE), again computed across the $50$ simulations and intervened nodes, under each method and for different values of $q$ and $n$.
While all methods improve as $n$ grows for both values of $q$, our Bayesian methodology based on a BMA estimate of the causal effect outperforms the IDA method under all scenarios. The accuracy of IDA is strictly related to the poor performance of the PC algorithm in recovering the true CPDAG. This in turn affects the correct identification of the set of distinct causal effects leading to the IDA estimate. By contrast, our BMA output is based in general on a larger collection of DAGs which, though possibly  outside the
equivalence class of the true CPDAG, may well lead to a causal effect which is closer to the nominal value because of structural ``causal" similarities with the true DAG.

\section{Application to anxiety and depression data}

We consider a dataset relative to a study on depression and anxiety in undergraduate students.
Depression represents a serious illness especially among young people which can be identified
through several symptoms such as feelings of melancholy and emptiness, disturbed sleep, or loss of interest in social activities.
In addition, it is strictly related to anxiety disorders and stress.
Several therapies for the treatment of depression and anxiety have been proposed and many of these have shown beneficial effects on patients in terms of a complete or partial restore of social behaviour and mental conditions.

The dataset, which is publicly available at
\url{https://www.kaggle.com/datasets/} under the name \textit{Depression and anxiety data},
was collected on $n=787$ undergraduate students from the University of Lahore.
Variables in the analyzed dataset include: depression diagnosis (the absence/presence of depressive status), anxiety diagnosis (\texttt{anx}, the absence/presence of anxiety disorder), and two related variables indicating the administration or not of a therapy against depression or anxiety (\texttt{depr treat} and \texttt{anx treat} respectively), besides other features such as \texttt{gender}, body max index (\texttt{bmi}, a categorical variable with two levels, normal/abnormal), suicidal instinct (\texttt{suicidal}),
and two variables linked to daytime sleepiness: \texttt{sleep} and its measure based on the Epworth scale (\texttt{epworth}).
Most variables are recorded as binary; scores were instead dichotomized.

\begin{figure}
	\begin{center}
		\includegraphics[scale=0.58]{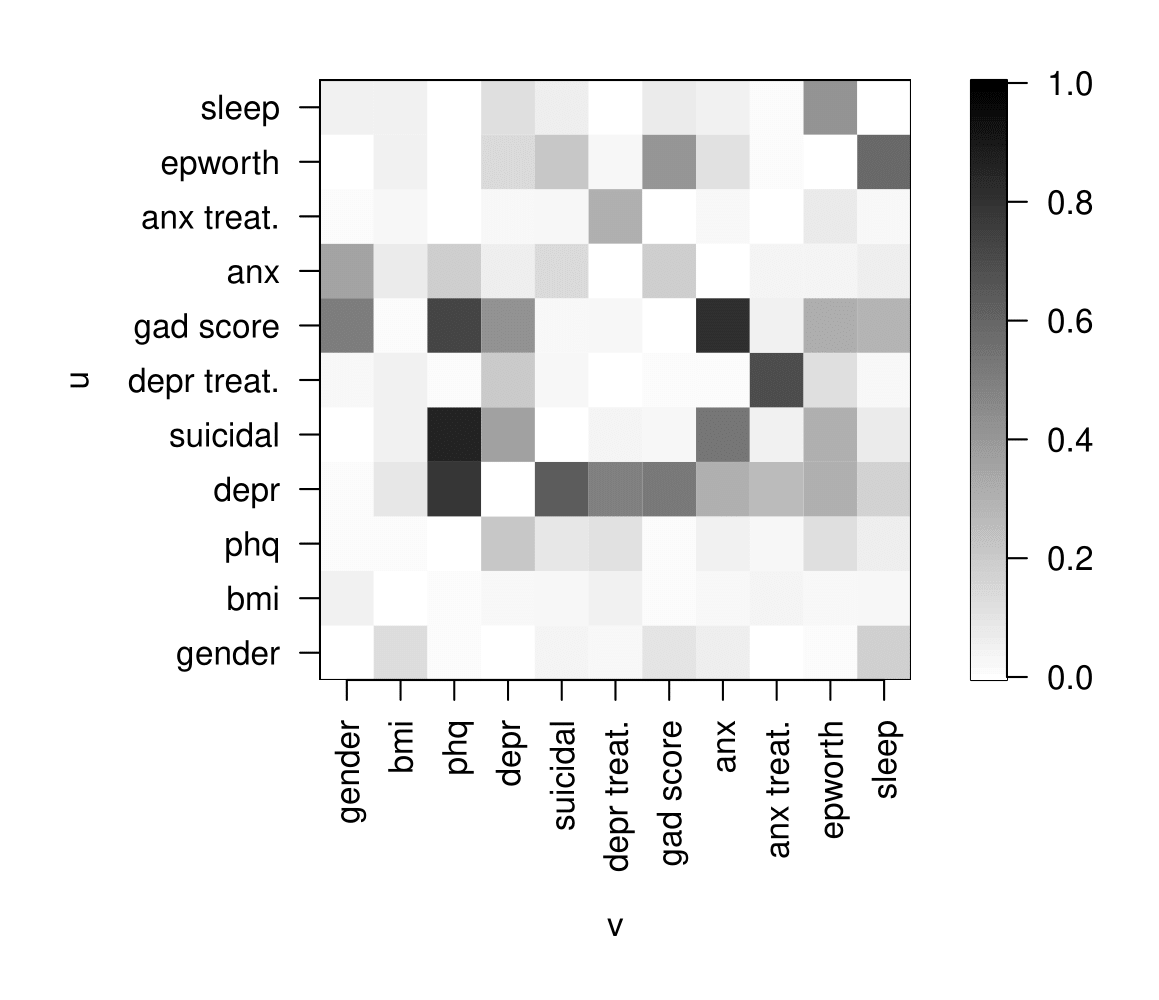}
		\includegraphics[scale=0.78]{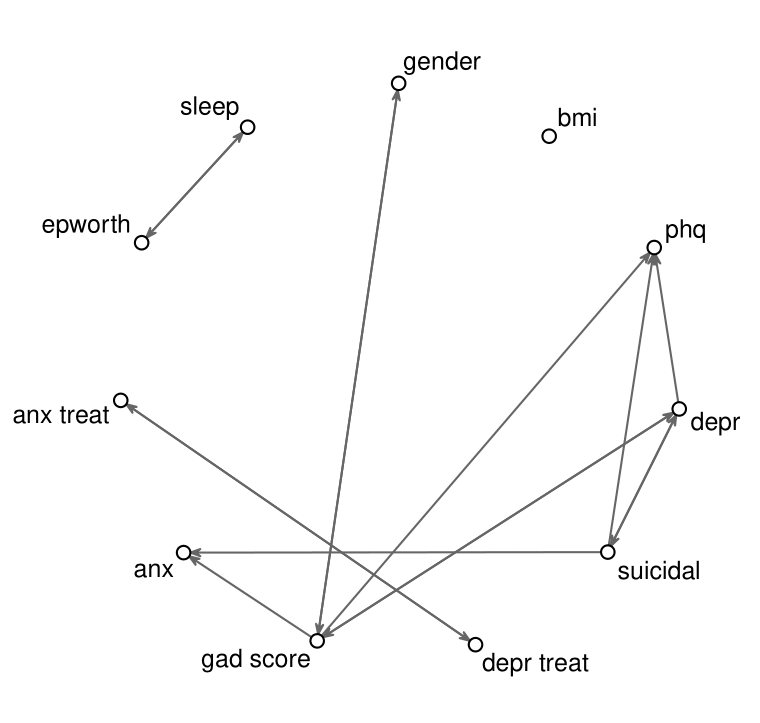}
		\caption{\small Anxiety and depression data. Heat map with estimated posterior probabilities of edge inclusion for each edge $u\rightarrow v$ (left panel); CPDAG representing the equivalence class of the median probability DAG estimate (right panel).}
		\label{fig:heatmap:dag:depression}
	\end{center}
\end{figure}

We implement our method for structure learning and causal effect estimation by running $S=40000$ iterations of our MCMC scheme after a burn-in period of $5000$ runs.
We summarize the output by reporting, for each directed edge $u\rightarrow v$ and each pair of variables in the dataset, the corresponding posterior probability of inclusion (Equation \eqref{eq:PPI}). Results are displayed in the heat map reported in the left-side panel of Figure \ref{fig:heatmap:dag:depression}.
In addition, we provide a summary of the posterior distribution over the DAG space by constructing the MPM DAG estimate. The CPDAG representing the Markov equivalence class of the estimated graph, which is reported in the right-side panel of Figure \ref{fig:heatmap:dag:depression}, is highly sparse as it contains only $10$ edges,
together with 3 unrelated components (in addition to the separate variable BMI): one involving the anxiety-depression diagnosis/measurement variables, one the two treatment variables, and finally the sleepiness block.

Variables which appear to be directly linked to  depression status are \texttt{phq} (Patient Health Questionnaire score) and \texttt{gad score} (Generalized Anxiety Disorder index), here included as binary variables with levels high and low, besides \texttt{suicidal}.
On the other hand, both \texttt{gender} and \texttt{bmi} do not seem to influence directly the depression or anxiety status.

%
%

\black
We now focus on causal effect estimation. Specifically, it is of interest to evaluate the efficacy of the two therapies for depression and anxiety.
Accordingly, we consider \texttt{depr} as the response of interest $Y$ in our causal-effect analysis and evaluate the causal effect onto \texttt{depr} of an intervention on \texttt{depr treat} $(X_v)$; similarly, we repeat the same analysis for intervention target \texttt{anx treat} and response variable \texttt{anx}.

\black
We recover from our MCMC output the posterior distribution of the two causal effect parameters  computed according to Equation \eqref{eq:gamma:v}. Summaries of the two resulting distributions, in terms of posterior mean, standard deviation and quantiles of order $(0.05,0.95)$ are reported in Table \ref{tab:depression}.
The posterior means of the two coefficients, corresponding to our BMA estimates, are both around $-0.12$ suggesting that both therapies have a beneficial effect on the status of depression and anxiety. In addition, both upper limits of the  $90\%$ credible intervals are close to the zero value, meaning that both distributions are much more concentrated on negative values, again supporting the result of a ``significant" (negative) causal effect on the two responses.

\begin{table}
	\centering
	\begin{tabular}{ccccc}
		\hline
		& Mean & St. Dev. & Quantile 0.05 & Quantile 0.95 \\
		\hline
		\multirow{1}{2.5cm}{Depression} & -0.117 & 0.243 & -0.685 & 0.010 \\
		\hline
		\multirow{1}{2.5cm}{Anxiety} & -0.126 & 0.222 & -0.649 & 0.022 \\
		\hline
	\end{tabular}
	\vspace{0.2cm}
	\caption{\label{tab:depression} \small Anxiety and depression data. Posterior summaries (mean, standard deviation and quantiles) for the two causal effect coefficients considered in the study.}
\end{table}


\section{Discussion}


In this paper we  consider
multivariate categorical observations
and
 propose a novel graphical model framework for causal inference.
  Specifically, our Bayesian methodology combines structure learning and parameter inference  for categorical Directed Acyclic Graph (DAG) models.
From a computational perspective, we implement a Markov Chain Monte Carlo (MCMC) scheme based on a Partial Analytic Structure (PAS) algorithm to approximate the joint posterior distribution over DAG structures and DAG parameters.
Starting from this MCMC output,
the full posterior distribution of the
causal effects between any pairs of variables of interest can be recovered, and eventually   summarized through  Bayesian Model Averaging (BMA), which naturally incorporates  uncertainty around the (unknown) underlying  DAG model.
We evaluate our method through simulation studies,  and demonstrate that it outperforms alternative state-of-the-art strategies
for causal effect estimation.
Additionally our method 
employs exact formulas based on conditional probabilities when computing  causal effects, and does not require further assumptions unlike in  \citet[Supplement]{Kalish:et:al:2010}.
\black

Our model formulation is based on the assumption of  i.i.d.~sample observations   from a \textit{single} categorical graphical model (which however is unknown,  or rather uncertain from a Bayesian perspective).
This assumption can be relaxed in two different directions to allow for
heterogeneity
among individuals belonging to different subgroups of the same population.
When groups are known beforehand, one can consider  
a model comprising  \textit{multiple} distinct graphical structures coupled with a Markov random field prior that encourages common edges between groups, and a spike-and-slab prior on network relatedness parameters \citep{Castelletti:et:al:2020:SIM}.
Causal effect estimation at group-specific level would benefit from  borrowing  information across subjects belonging to distinct, yet related groups.

On the other hand, when subgroups are not available \textit{a priori}, one can set up a \textit{mixture} model, either with a finite or an  infinite number of components, allowing for
joint posterior inference on DAGs, parameters as well as clustering.
A Bayesian non-parametric Dirichlet Process  mixture of Gaussian DAG-models is considered in
\citet{Castelletti:Consonni:2023:SIM}
for causal inference under heterogeneity.
Their general framework can be adapted to categorical DAGs and would lead to causal effect estimates  at cluster as well as subject-specific level.


\bibliographystyle{Chicago}
\bibliography{biblio}

\end{document}